# Observation of Huge Magnetoresistance and Multiferroic-like Behavior of Co Nanoparticles in a $C_{60}$ matrix

Yutaka Sakai * [1,2], Eiiti Tamura * [2,#], Shuhei Toyokawa [1], Eiji Shikoh [1],
Vlado K. Lazarov [3], Atsufumi Hirohata [4,5],
Teruya Shinjo [1,2], Yoshishige Suzuki [2] and Masashi Shiraishi [1,5,#]

1. Department of Systems Innovation, Graduate School of Engineering Science, Osaka University. 560-8531 Toyonaka, Japan.
2. Department of Materials Engineering Science, Graduate School of Engineering Science, Osaka University. 560-8531 Toyonaka, Japan.
3. Department of Physics, University of York, Heslington, York, YO10 5DD, United Kingdom.
4. Department of Electronics, University of York, Heslington, York, YO10 5DD, United Kingdom.
5. PRESTO-JST, 4-1-8 Honcho, Kawaguchi 332-0012, Saitama, Japan.

* These two authors contributed equally to this work.
# Corresponding authors. shiraishi@ee.es.osaka-u.ac.jp, tamura@spin.mp.es.osaka-u.ac.jp

**Tunneling magnetoresistance (TMR) [1,2] via oxides or molecules includes fruitful physics, such as spin filtering [3] and hybridized interface states [4], in addition to various practical applications using large TMR ratio at room temperature [5]. Then, a larger TMR effect with a new fundamental physics is awaited because further progress on spintronics can be realized. Here we report a discovery of a gigantic TMR ratio of 1,400,000% in a $C_{60}$-Co nanocomposite spin device. The observed effect is induced by a combination of a Coulomb blockade effect and a novel magnetic switching effect. Theoretical investigation reveals that an electric field and a magnetic field control the magnetization and the electronic charging state, respectively, of the Co nanoparticles as in physics of multiferroics.**

Molecular tunnel barriers were originally intended as a simple replacement of insulating barriers such as AlO or MgO, several unique features were observed in both stacked TMR devices and nano-composite (granular) devices. For example, recent studies demonstrated (1) a very large TMR at low temperatures [4] and spin-dependent tunneling transport at room temperature [6,7], (2) the existence of a higher-order (~5th) co-tunneling effect [8], and (3) an enhancement of the spin polarization of ferromagnets at the interface between the ferromagnets and molecules [9,10]. Therefore, it is now recognized that molecular TMR involves novel physical aspects which have not been observed in metallic or inorganic spintronics. Among these aspects, the large magnetoresistance (MR) ratio of 300% at 2 K observed by Barraud and co-workers [4] represents a new frontier in molecular spintronics, because the origin of this large MR was clarified, enabling new options in device design through the incorporation of molecules. However, stronger effects and/or novel physical aspects are in strong demand for further progress of spintronics including molecular spintronics. Here, we report on huge MR of 1,400,000% appeared in a $C_{60}$-Co nanocomposite spin device, in which ferromagnetic Co nanoparticles were uniformly dispersed in a molecular matrix and the $C_{60}$ behaved as a tunneling barrier [6-11]. A theoretical modeling, calculations and supporting experiments manifest its physics to be unprecedented multiferroic-like behavior of the Co nanoparticles.

Figure 1a shows a schematic of the device structure, in which the gap length, $L$, was varied from 1.5 to 15 μm (see the Methods section). The compositional ratio of $C_{60}$:Co was estimated to be ~ 9:1 according to the co-evaporation rates. As shown in Fig. 1b, the Co nanoparticles were roughly spherical and uniformly dispersed in the $C_{60}$ matrix, and their mean diameter was estimated to be 2.5 ± 0.4 nm on the basis of transmission electron microscopy (TEM) observations. It was also clarified that Co-nanoparticles and $C_{60}$ matrix were packed without any visible defects, suggesting the samples were free from any possible magnetostriction or Co-particle motion by bias voltage applications. A characteristic feature appeared in the *I-V* curves, namely, there was an obvious discontinuity in the *I-V* curve at +7.83 V under zero magnetic field (see Fig. 1c). This discontinuity was due to a Coulomb blockade effect, judging from the change in the *I-V* curves with increasing temperature. The discontinuity disappeared as the temperature increased, and non-linearity was only observed in the curve above 30 K. It is noteworthy that we previously ascribed the Coulomb blockade to a charging effect in the Co nanoparticles [11]. As shown in Fig. 1d, the threshold voltage at 2 K increased linearly as the gap length increased, which demonstrates the uniformity of our samples.

Figure 2a shows the magnetic field dependence of the *I-V* curves observed in a device of *L*=5 μm at 2 K, for magnetic fields of up to 5 T. It should be emphasized that (1) the threshold voltage exhibited a dependence upon the external magnetic field, dropping to +7.6 V at 5 T, and (2) the shift was saturated once the magnetic field reached 5 T. Therefore, our finding was due to a magnetism-induced effect, and a magnetoresistance effect was revealed. It was previously reported that the magnetoresistance effect in such a device is governed by a relative angle of magnetization in the Co particles (see, for example, ref. 7). Therefore, it is likely that the saturation of the voltage shift corresponds to a saturation of the magnetization of the Co nanoparticles. In other words, the saturation corresponds to a change in the tunnel conductance between the Co nanoparticles below the threshold voltage. Fig. 2b shows the correspondence between the sample resistance below the threshold voltage and the shift of the threshold voltage. The normalized values of both quantities were in good accordance, verifying the above argument and demonstrating that the observed magnetoresistance effect was not spurious.

A shift of the threshold voltage was also observed during a backward sweep of the bias voltage, which would not be expected in a conventional spin device of similar structure (see the inset of Fig. 2a). Here, we note that the threshold voltages in the hysteresis were shifted under the application of an external magnetic field, and also that the shifting voltage under a forward sweep was larger than that under the backward sweep. The threshold voltages in both the forward and backward sweeps downshifted under an applied magnetic field; that is, the resistance of the device was decreased by the application of a magnetic field. Therefore, the finding was not caused by a spin blockade effect, because the tendency of the dependence of the resistance on the magnetic field was reversed. Nor could the finding be ascribed to electrical breakdown, because the obtained results were reproducible and repeatable.

The appearance of the magnetoresistance effect enabled an estimation of MR ratios of 400,000% and 1,400,000% during forward and backward sweeps, respectively (see Fig. 2c). Spin motive force [12] (the magnetoresistance ratio was ~ 100,000% at 2 K) might be a plausible origin of this large effect. However, it should be emphasized that there was no shift of the *I-V* curves around a zero bias voltage with or without an external magnetic field, which eliminates this possibility. It is possible that our definition of the MR ratio was somehow exaggerated, because the electric current below the threshold voltage was strongly suppressed. Therefore, this phenomenon should be called a novel magnetic switching effect because the spin alignment (the magnetization direction) of the Co nanoparticles was switched by an external *electric* field. (The detailed mechanism of this switching effect is discussed in the following paragraphs.) The on/off ratio of this magnetic switching device is calculated to be $4.0 \times 10^3$ and $1.4 \times 10^4$. Other evidence supports the conclusion that our finding is due to a switching effect. The switching behavior in this study was observed when *the bias voltage* was swept in a fixed *external magnetic field* in previous experiments. If this phenomenon was driven by a switching of the magnetic alignments of Co nanoparticles, a similar switching in the resistance should be observed in a changing *external magnetic field* with a fixed *bias voltage*. Figure 3 shows an example of magnetic-field-induced switching in the same device that exhibited the bias-voltage-induced switching. When we fixed the bias voltage at +7.63 V, which was the threshold voltage at 3 T and was between the threshold voltages at 0 T and 5 T, the resistance of the device changed dramatically (by more than 3 orders of magnitude) at 3 T during the forward sweep of the magnetic field, whereas no switching was observed in the backward sweep as was expected from the hysteresis in the *I-V* curves. From above results, it is clarified that the Co nanoparticles exhibited multiferroic-like behavior because magnetoelectric coupling was observed.

Figure 4 shows the experimental (upper panels) and theoretical (lower panels) magnetization curves at temperatures well above the blocking temperature $T_B$ which was determined by the magnetic susceptibility measurements. The blocking temperature $T_B$ for the 6:1 composition was estimated at a peak temperature 16 K of the zero field cooling susceptibility (inset of the upper left panel), and $T_B$ for the 9:1 composition was 10K (not shown). The experimental curves are plotted as a function of the reduced external magnetic field $|\mathbf{H}|/T$ at temperatures of 20 K, 35 K and 50 K. If the magnetic moments of Co nanoparticles are magnetic-anisotropy-free and interaction-free to the other particles, then all different temperature curves should be on a universal curve so-called the Langevin function (the black thin lines in lower panels). The observed magnetization curves deviate strongly from the universal curve, and have a common feature that is a steep increase of the magnetization until a half of the saturation magnetization $M_S$, while 80% of $M_S$ for the universal Langevin function. The feature is typical for a magnet with a uniaxial magnetic anisotropy. Assuming that the uniaxial direction (the magnetic easy axis), to which the magnetic moments of the nanoparticles stick, is inclined at an angle of $\theta$ from the magnetic field $\mathbf{H}$, the average magnetization with respect to the angle $\theta$ is given by

$$\langle \mathbf{M} \rangle = \langle M_Z \rangle = M_S \int_0^{\pi} d\theta \rho(\theta) |\cos\theta| = \frac{1}{2} M_S \, , \qquad (1)$$

where the weight $\rho(\theta) = (1/2)\sin\theta$ is given in a condition such that $\int_0^{2\pi} d\phi \int_0^{\pi} d\theta \sin\theta / 4\pi = \int_0^{\pi} d\theta \rho(\theta) = 1$. To obtain a larger magnetization than $M_S/2$, the magnetic moments has to move away from the magnetic easy axis at the price of the energy. To analyze the magnetic structures further, we have performed numerical simulations to several magnetic models with and without interactions between particles. Since the particles are randomly and sparsely distributed, only the interaction between nearest neighboring particle pair is taken into account for calculating the Maxwell-Boltzmann statistical weight, $Z(|\mathbf{H}|,\theta) = \iint d\Omega_1 d\Omega_2 \exp[-H(\mathbf{m}_1,\mathbf{m}_2;\mathbf{H})/k_B T]$. Among the models including the Heisenberg model, the following two models are consistent with the experimental magnetization curves. They are (i) non-interacting particle model with a uniaxial magnetic anisotropy and (ii) magnetic dipolar interacting two-particle model (Fig.5a) and their Hamiltonians are given as

$$H_A = -\mathbf{m} \cdot \mathbf{H} + \frac{K}{m}\left\{m^2 - (\mathbf{m} \cdot \hat{\mathbf{r}})^2\right\} \quad , \tag{2}$$

$$H_D = -(\mathbf{m}_1 + \mathbf{m}_2) \cdot \mathbf{H} + J\left\{\mathbf{m}_1 \cdot \mathbf{m}_2 - 3(\mathbf{m}_1 \cdot \hat{\mathbf{r}})(\mathbf{m}_2 \cdot \hat{\mathbf{r}})\right\} \; ; \; (J \geq 0) \quad . \tag{3}$$

We define the magnetic anisotropy energy $K$ *per* the unit magnetization so that $K$ is described in terms of the strength of magnetic fields. Since the shapes of Co nanoparticles are spherical, the anisotropy energy $K$ is exclusively attributed to the crystal-structural origin in the model (i). The coordinate setup of two neighboring particles for the Hamiltonian $H_D$ is shown in Fig.5a and the vector is defined as $\hat{\mathbf{r}} \equiv \mathbf{r}/|\mathbf{r}|$. The dipolar coupling constant $J$ is proportional to $|\mathbf{r}|^{-3}$ but we leave it as a fitting parameter in our model. After the integration over the solid angles of the magnetic moments and the average out for the angle $\theta$ with the weight $\rho(\theta)$, we obtain the statistical weight $Z(\alpha,\beta)$ in terms of the parameters $\alpha = M_S|\mathbf{H}|/k_B T$ and for the models (i) $\beta = M_S K/k_B T$ and (ii) $\beta = M_S^2 J/k_B T$ , respectively. The normalized magnetizations are given by $\langle \mathbf{M} \rangle = \frac{\partial}{\partial \alpha} \ln Z(\alpha,\beta)$. We show the numerical results for the model (ii) in Fig.4. The moments are tend to be parallel through the dipolar interactions, as is seen <cos γ>=0.8 for the zero external field at $T$=20 K (inset of the lower panels) and the two models (i) and (ii) are equivalent when $\mathbf{m}_1 = \mathbf{m}_2$. Two Co particles act as a single magnetic domain particle with a uniaxial magnetic anisotropy (Fig. 5b). As we have seen that the shape of the magnetization curves are predominantly determined by the kinks at $M_S/2$ which are moderate in the experimental curves due to the distribution of the particle size and of the distance between particles. The both effects are neglected in our theoretical simulations. The estimated anisotropy energies $K$ are 3500(Oe) for the compositional ratios of $C_{60}$:Co of 6:1 and 26000(Oe) for 9:1 while 3100(Oe) for the Co hcp crystal. The latter value of 26000(Oe) is about 7.5 times larger than the former value, that cannot be attributed only to the difference of the mean particle size due to the compositional ratios. Although neither models (i) nor (ii) are excluded from the magnetic structure of the present Co-$C_{60}$ systems and also their coexistence is likely, the anomalously enhanced anisotropy energies for the ratios of $C_{60}$:Co of 9:1 could be attributed to the shape anisotropy produced by the two Co particle pairs rather than the crystal anisotropy energy only.

In our previous study [11], it was clarified that (1) the appearance of a threshold voltage due to a Coulomb blockade effect in a molecular nanocomposite was caused by co-tunneling via the Co nanoparticles, (2) there were several bottleneck structures at which the electric field and the electric current were concentrated, and (3) the magnetization alignment of the Co nanoparticles in the bottleneck structure governs the spin transport properties. Another key to explaining the present observations is the hysteresis curves shown in Fig. 2a. The area inside the hysteresis loop equals the work done in a cycle, which will be dissipated eventually as heat. The hysteresis loop without a magnetic field encloses a larger area than the loops with a magnetic field. In the latter case, especially the case for the saturated magnetic field of $H_s$=5 T, we can assume that all of the magnetic freedoms are frozen. Therefore, we must attribute the dissipation for the hysteresis to a nonmagnetic origin. Furthermore, magnetic dissipation plays an important role when no magnetic field is present or in weaker magnetic fields. Although the substrate temperature varied the hysteresis with the threshold position (Fig. 1c), repeated observations, which may have increased the temperature, did not alter the hysteresis curves. This can be understood as follows: The current after the breakdown of the Coulomb blockade is not dissipative at the bottleneck nanoparticles (ballistic conduction) and does not raise the local temperature, which is responsible for the repeatedly observed hystereses. Instead, the current is dissipated in the surroundings, increasing the substrate temperature. On this basis and our magnetic structure analysis in the previous paragraph, we constructed a simple two-nanoparticle model for the qualitative analysis of this novel magnetic switching effect (see Fig. 6a). In the model, the two particles form a bottleneck structure for electron transport, and the magnetic alignment of the two ferromagnetic particles determines the conductance of the system. We employed two assumptions. The first assumption is that the magnetization of the two Co nanoparticles under zero magnetic field is aligned anti-parallel due to magnetic dipolar interaction (Fig. 5c). The second assumption is that the two particles act as a single bottleneck because the particles are positioned in parallel to the surrounding contacts, rather than in series. As shown in Fig. 6a, the wavefunction of the injected electron is localized in one particle when the magnetic alignment is anti-parallel, because of the symmetry of the wave function. On the other hand, the wavefunction is distributed over both particles when the magnetic alignment is parallel. In the latter case, the charging energy of the system was smaller than in the former case; that is, the system is stable in the parallel magnetic configuration, when the electron can move around over the two particles. It is worth noting that ferromagnetic nanoparticles with diameters smaller than ~20 nm favorably form a single-domain magnetic structure for the same reason. The energy states of the magnetic configuration of the two Co particles cause the magnetic dissipation. On the other hand, for nonmagnetic dissipation, a polarization effect between the nanoparticles and the surrounding fullerene molecules should be taken into account when a single electron is injected into a Co nanoparticle or a Co-nanoparticle pair. The effect induces a decrease in the charging energy of the nanoparticle in which the electron is injected, yielding a decrease of the threshold voltage of the Coulomb blockade (Fig. 6b). In addition, it is noteworthy that fullerene has a small dielectric constant (~3). Therefore, the increase in charging energy caused by a single charge injection is inversely proportional to the dielectric constant, and is much larger than in other insulating granular systems ($SiO_2$, Al-O, etc.).

Figure 6c shows a schematic overview of the magnetic switching mechanism. We begin with the forward biased case without an applied external magnetic field; that is, with anti-parallel coupling of the Co nanoparticles. When the applied bias voltage is increased to reach the threshold voltage, one electron is injected into a Co nanoparticle with the charging energy $E_c$ as the Coulomb blockade is broken. The excited level of the charged state is high enough that the level could decay to a lower level corresponding to the parallel magnetization alignment of the two Co nanoparticles. As discussed above, the charging energy dropped to a smaller value $E_c$', and the threshold voltage in the backward sweep is additionally reduced by the polarization effect ($E_{cp}$). The appearance of hysteresis in the forward and backward sweeps without an external magnetic field can be ascribed to the above mechanism. When an external magnetic field of $H=H$s is applied, the basis state of the system was parallel (the blue state shown in Fig. 6c) and the magnetization of the system was frozen. The threshold voltage in the forward sweep is already shifted downward because of the parallel alignment, and that of the backward sweep was reduced because of the magnetic dipolar interaction. The interaction energy was much smaller than $E_c$' and $E_{cp}$, so the threshold voltage shift in the downward sweep was smaller than of the forward sweep. In our model, the threshold voltage in the forward sweep is governed by the charging energy, which is a function of the magnetic alignment of the two ferromagnetic nanoparticles, and the co-tunneling current below the threshold is also dependent on the magnetic alignment. The experimental results shown in Fig. 2b are consistent with the model. The observed hysteresis can be ascribed to dissipation of the charging energy. This dissipated energy induces magnetic switching, and the hysteresis curves in the forward and backward sweeps shift differently with respect to a external magnetic field. If no magnetic switching occurred, so that the magnetic alignment was fixed during the bias-voltage sweep, the areas within the hysteresis loops would be independent of the magnetic field, and the loops would merely shift as the thresholds shifted (See Fig. 6c). It is noteworthy that our asymmetric two-particle model, representing the model in Fig.5c, would be preferable to a symmetric model for magnetic decay from the anti-parallel to the parallel states, because in the latter case neither nanoparticle is able to absorb a reaction against the spontaneous parallel alignment, while one of two particles is pinned to the local field in the former case. Hence, the experimentally observed shift in the hysteresis during forward and backward sweeps clearly indicates magnetization switching, in which a charging state controls magnetization and a magnetic field controls a charging state, resembling a multiferroic effect. The first assumption of our model that the two Co nanoparticles under zero magnetic field is aligned anti-parallel, seems peculiar, since magnetic dipolar interactions induce not only anti-parallel coupling but also many others, including parallel coupling. But the other coupling cannot be a bottleneck, because their charging energy is lower than that of the anti-parallel, and therefore cannot be observed in the present experiment. On the other hand, a very isolated Co nanoparticle cannot take part of a conducting network, although it may have higher charging energy enough to be a bottleneck. It seems to be a phenomenon as if the experiment had chosen a chance to make a field-sensitive Coulomb-blockade network. We have seen that the multiferroic-like behavior such as the voltage-control magnetization can be elaborated in terms of artificial devices rather than material objects.

The use of this model to explain the experimental observations was merely qualitative, because the actual device consisted of several bottleneck structures, judging from the comparably large electric current after the Coulomb blockade was broken. Although modification of the device structures and improvements in the modeling are required for a quantitative understanding of this novel magnetic switching effect, it should be emphasized that the present simple qualitative model does explain the basic mechanism of the switching effect. It is worthy to note that the phenomena are strongly dependent on the composition ratios of $C_{60}$-Co and these interesting effects show up in the case of the ratio close to 9:1 where the magnetic anisotropy energy is anomalously large and it could not be attributed to the crystal anisotropy energy of a single Co particle. Currently, this switching effect disappears at ~20 K because it is a Coulomb-blockade-induced effect. However, controlling the diameter of the ferromagnetic nanoparticles may allow an increase in the temperature at which the effect appears. We note that this effect may be observed in other matrix materials, such as Al-O or $SiO_2$. However, the introduction of molecular materials with a small dielectric constant is a key to inducing this effect, because the charging energies of the basic and excited states are widely separated. The introduction of molecules is a comparably new approach, and unknown issues remain. Detailed investigations should be vigorously pursued in the future to obtain stronger effects at higher temperatures.

## Methods

### Sample fabrication

A $C_{60}$-Co nanocomposite was fabricated on an $Si/SiO_2$ substrate with Au (40 nm)/Cr (3 nm) electrodes by a co-evaporation method. The channel length between the electrodes was varied from 1.5 μm to 15 μm. The purity of the $C_{60}$ was 99.99%. The substrate temperature during the co-evaporation was an ambient temperature. The composition ratio of $C_{60}$ : Co was 9 : 1, and was controlled by the growth rates of both materials (~0.9 A/s for $C_{60}$ and ~0.1 A/s for Co). After evaporating the $C_{60}$-Co to 150 nm, capping layers of $C_{60}$ and $SiO_2$ (300 nm and 240 nm thick, respectively) were evaporated continuously in order to prevent oxidation of the Co nanoparticles. The samples for TEM observation were prepared individually, and had a composition ratio of 8.6:1.

**Sample characterization**

*I-V* curves were measured using a Physical Property Measurement System (PPMS, Quantum Design Co.) and a source meter (Keithley 2400), with an external magnetic field of up to 5 T applied perpendicular to the $C_{60}$-Co film. The temperature was varied from 2 K to 100 K. In our previous studies, we reported that the magnetization of Co nanoparticles induced a magnetic field dependence of the electric current, which was called the magnetoresistance effect [10], and that a Coulomb blockade occurred in the nanocomposite films under investigation [11]. The MR ratio was defined as 100 × {*I* (*B*=5 T)–*I* (*B*=0 T)}/ *I* (*B*=0 T). TEM images were acquired on a double aberration corrected JEOL 2200FS high resolution field emission transmission electron microscope, operated at 200 keV. Cross-sectional TEM specimens were prepared using conventional methods, which included mechanical thinning and polishing followed by Ar ion beam milling to achieve specimen electron transparency. The mean diameter of the Co nanoparticles was estimated by averaging the diameter values of one hundred nanoparticles.

# References

1. Miyazaki, T. & Tezuka, N. Giant magnetic tunneling effect in Fe/Al2O3/Fe junction. *J. Mag. Mag. Mat.* **139**, L231-L234 (1995).

2. Moodera, J.S., Kinder, L.R., Wong. M. & Meservey, R. Large magnetoresistance at room temperature in ferromagnetic thin film tunnel junctions. *Phys. Rev. Lett.* **74**, 3273-3276 (1995).

3. Butler, W.H., Zhang, X.-G., Schulthess, T.C., and MacLaren, J.M. Spin-dependent tunneling conductance of Fe|MgO|Fe sandwiches. Phys. Rev. B **63**, 054416 (2001).

4. Barraud, C et al., Unraveling the role of the interface for spin injection into organic semiconductors. *Nature Phys.* **6**, 615 (2010).

5. Yuasa, S., Nagahama T., Fukushima, A., Suzuki Y. & Ando K. Giant room-temperature magnetoresistance in single-crystal Fe/MgO/Fe magnetic tunnel junctions. *Nature Mat.* **3**, 868-871 (2004). Also, Parkin, S.S.P., Kaiser, C., Panchula, A., Rice, P.M., Hughes, B., Samant, M. & Yang, S-H. Giant tunneling magnetoresistance at room temperature with MgO (100) tunnel barriers. *Nature Mat.* **3**, 862-867 (2004).

6. Shim, J.H., Raman, K.V., Park, Y.J., Santos, T.S., Miao, G.X., Satpati, B & Moodera, J.S. Large Spin Diffusion Length in an Amorphous Organic Semiconductor. *Phys. Rev. Lett.* **100**, 226603 (2008).

7. Miwa, S., Shiraishi, M., Mizuguchi, M., Shinjo T. & Suzuki Y. Spin-dependent transport in $C_{60}$-Co nano-composites. *Jpn. J. Appl. Phys.* **45**, L717-L719 (2006).

8. Hatanaka, D. et al. Enhanced magnetoresistance due to charging effects in a molecular nanocomposite spin device. *Phys. Rev. B* **79**, 235402 (2009).

9. Shiraishi, M., Kusai, H., Nouchi, R., Nozaki, T., Shinjo, T., Suzuki, Y., Yoshida M. & Takigawa M. A nuclear magnetoresistance study on rubrene-cobalt nanocomposites. *Appl. Phys. Lett.* **93**, 053103 (2008).

10. Matsumoto, Y. et al. X-ray absorption spectroscopy and magnetic circular dichroism in codeposited $C_{60}$–Co films with giant tunnel magnetoresistance. *Chem. Phys. Lett.* **470**, 244-248 (2009).

11. Miwa S., Shiraishi, M., Tanabe, S., Mizuguchi, M., Shinjo, T. & Suzuki. Y. Tunnel magnetoresistance of $C_{60}$-Co and spin-dependence transport in organic semiconductor. *Phys. Rev. B* **76**, 214414 (2007).

12. Hai, P.N., Ohya, S., Tanaka, M., Barnes, S.E. & Maekawa, S. Electromotive force and huge magnetoresistance in magnetic tunnel junctions. *Nature* **458**, 489-492 (2009).

**Figure legends**

**Figure 1 | Device structures and electrical characterizations of the $C_{60}$-Co nanocomposite spin devices. a.** A schematic of the $C_{60}$-Co nanocomposite spin device. The nanocomposite film was evaporated onto a $SiO_2$/Si substrate. The Co nanoparticles were uniformly dispersed in the $C_{60}$ matrix, and the nanocomposite film was covered by a $C_{60}$ film 300 nm thick and a $SiO_2$ film 240 nm thick to prevent oxidation of the Co nanoparticles. The electrodes were Au/Cr (40/3 nm). The external magnetic field was applied perpendicular to the film. The channel length $L$ was varied from 1.5 to 15 μm. **b.** (Top) A TEM images of a $C_{60}$-Co nanocomposite film. The $C_{60}$-Co nanocomposite film is seen on the $SiO_2$ film. Black regions correspond to the Co nanoparticles, the average diameter of which was typically 2.5 nm. (Bottom) An enlarged view of the $C_{60}$-Co nanocomposite film. **c.** A temperature evolution of the *I-V* curves observed in the $C_{60}$-Co nano-composite device ($L$=5 μm) without the application of an external magnetic field. An apparent discontinuity in the *I-V* curves can be seen at up to 20 K due to a Coulomb blockade effect. The threshold voltages of the Coulomb blockade increased from 5.3 V to 7.8 V as the temperature decreased. The non-linear *I-V* curve disappeared above 70 K, which was the upper limit of the appearance of the Coulomb blockade. Other discontinuities in the *I-V* curves at higher bias voltages (from 6.5 V to 8.0 V) can be seen at up to 20 K, which also indicates that this characteristic feature was attributed to the Coulomb blockade. **d.** Channel length dependence of the threshold voltages at 2 K in the $C_{60}$-Co nano-composite devices. The threshold voltage increased linearly as a function of the channel length, further evidence of a Coulomb blockade effect.

**Figure 2 | Magnetic switching and magnetoresistance effects in the $C_{60}$-Co nanocomposite. a.** Magnetic field dependence of the *I-V* curves and the appearance of hysteresis. The dashed arrows show the directions of bias voltage sweep. Apparent hysteresis was observed in the forward and

backward bias sweeps. The inset shows the shift of the threshold voltage in the backward sweeps under 0 and 5 T fields. **b.** Correspondence between the threshold voltage and sample resistance at a bias voltage of 6.5 V. The vertical axis shows a normalized value of the sample resistance at 6.5 V (a black solid line) and the threshold voltages under various magnetic fields (colored open circles). This normalization was implemented by values of the threshold voltages and resistance at 5 T, as shown in the figure, where $A$ indicates a physical parameter (the threshold voltage or the resistance). The colors of the open circles correspond to that of the $I$-$V$ curves under the magnetic fields, as shown in Fig. 2a. Both values have good accordance, which directly indicates the existence of the Coulomb blockade effect. **c.** The MR ratio of the device ($L$=5 μm) at 2 K. The MR ratio was defined as $100 \times \{I(B=5\text{ T}) - I(B=0\text{ T})\}/I(B=0\text{ T})$. The MR ratio was calculated to be ca. 400,000% in the forward biasing (fw, a black solid line) and ca. 1,400,000% in the backward biasing (bw, a red solid line).

**Figure 3 | A magnetic-field-induced switching effect observed at 2 T in the $C_{60}$-Co nanocomposite.** The bias voltage was fixed at 7.85 V and the magnetic field was swept from 0 T to 5 T (forward biasing; black closed circles) and from 5 T to 0 T (backward biasing; red closed circles). At ca. 3 T, an obvious switching was observed in the forward biasing.

**Figure 4 | Experimental and theoretical magnetization curves.** The upper panels are experimental magnetization curves with respect to the reduced external field strength ($|\mathbf{H}|/T$) and the lower panels their theoretical curves, and the left and right columns are for the compositional ratios of $C_{60}$:Co of 6:1 and 9:1, respectively. The colors indicate different observation temperatures. The blocking temperature for the 6:1 composition was found to be 16K in the zero field cooling (ZFC) and the field cooling (FC) susceptibility (inset of the upper left panel). The theoretical

parameters, $\alpha = mH/k_BT$ and $\beta = m^2J/k_BT$, are fixed at the temperature of $T=35$K in comparison with the corresponding experimental data. The relative angles of the magnetic moments of the two particles are shown in the insets of lower panels. The angles tend to be parallel even without the external field due to the magnetic dipolar interaction. The black thin lines display the case in the absence of an interaction between the two particles, which are given by the Langevin function.

**Figure 5 | Magnetic and structural models of Co nanoparticles.** **a.** The coordinate setup of two neighboring nanoparticle for the model Hamiltonian. The relative position vector **r** is chosen in the z-axis and the external magnetic field **H** in the (x,z)-plane. **b.** The most stable magnetic configuration of the dipolar interacting two Co particles. When the external filed **H** is weak, the two Co particles act as a single magnetic domain particle with a uniaxial magnetic anisotropy. **c.** An antiferromagnetically coupled nanoparticle pair. When the particles align, their magnetic moments are easily pinned by each other through their dipolar interactions, and form a possible antiferromagnetic coupling with a horizontally located particle.

**Figure 6 | Theoretical modeling of the magnetic switching effect.** **a.** Schematic of the two-nanoparticle Coulomb blockade model. Although the anti-parallel magnetic configuration is energetically favorable in the discharged state, the parallel configuration becomes more favorable than the anti-parallel in the charged state. In the parallel configuration, the wavefunction of the injected electron can extend over both particles to reduce the charging energy $E_c$. The magnetic moment of the larger particle is pinned to the *local* magnetic field (cf. Fig.5c), while that of the smaller particle is sensitive to both the local field and the dipole field produced by the larger particle. **b.** Nonmagnetic contribution to the hysteresis in a charging-discharging process. To charge a nanoparticle, the bias voltage must overcome the charging energy $E_c$. Once the particle is charged,

electrostatic charge polarization of surrounding media reduces the charging energy by an amount $E_{cp}$, and a lower bias voltage is required to maintain the charged state. **c.** Schematic energy diagram of two nanoparticles. In the forward sweep, the external magnetic fields vary the threshold voltages with the charging energies $E_c(E_c')$, which depend on the magnetic configuration of the particles. Once the particles are charged (or once the Coulomb blockade is broken), the energy state of the particles drops to the energetically lowest charged state by exciting magnons and phonons so that the magnetic configuration is aligned in parallel and the surrounding media are fully polarized to reduce the charging energy. In the backward sweep, the discharging threshold voltages reflect the energy differences of the magnetic dipolar interaction $\Delta E_{dip}$ of two nanoparticles in the parallel magnetic configuration. With the saturation field, the parallel configuration is energetically more favorable by a similar amount of $\Delta E_{dip}$ than the antiparallel.

Figures

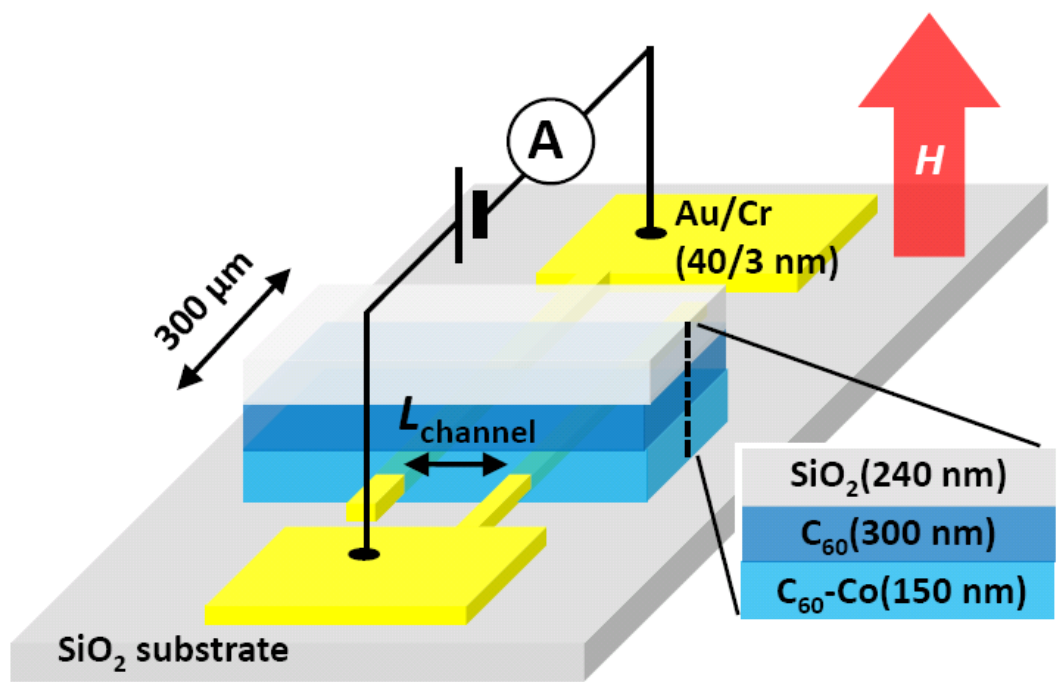

Fig. 1a Y. Sakai et al.

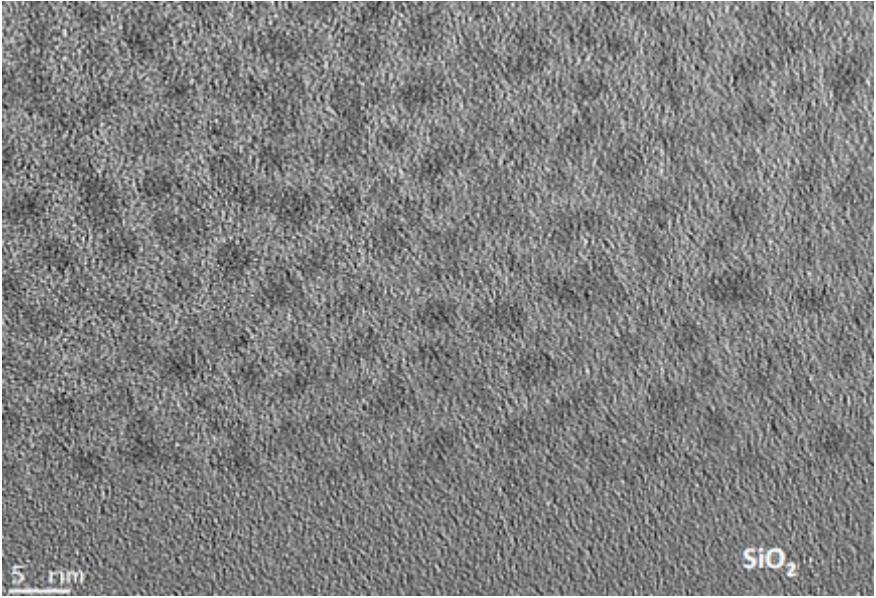
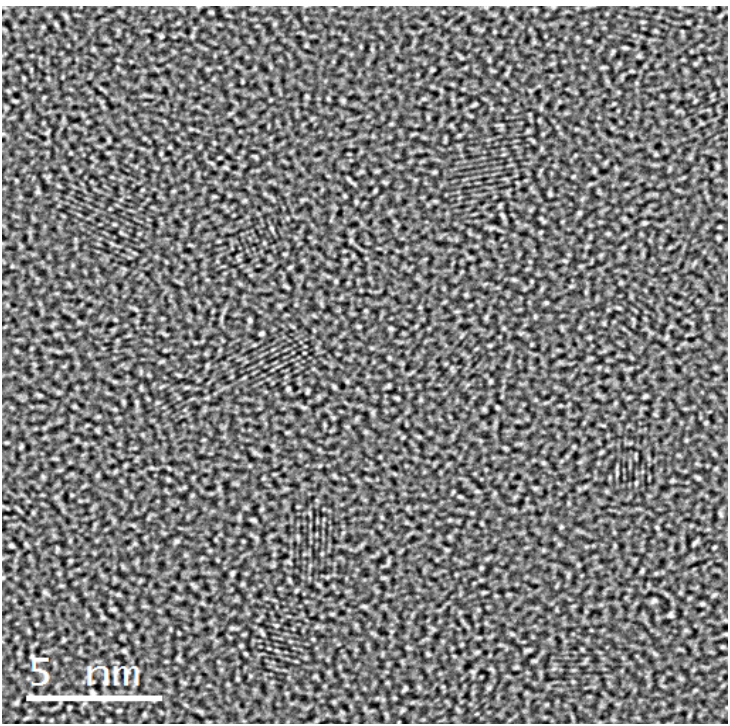

Fig. 1b Y. Sakai et al.

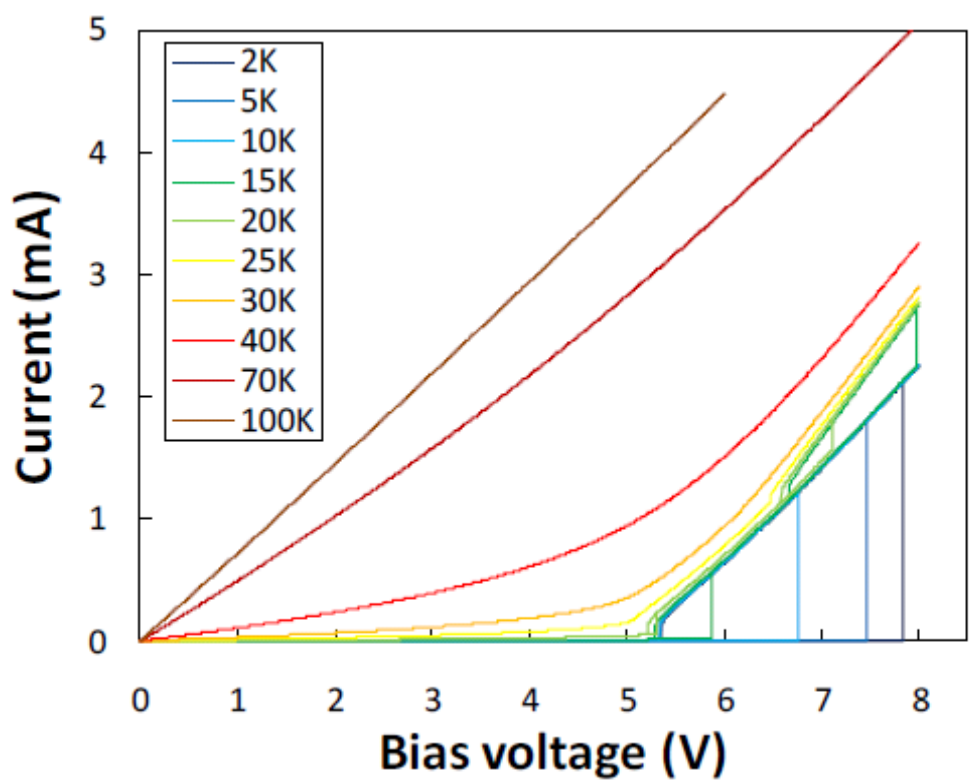

Fig. 1c Y. Sakai et al.

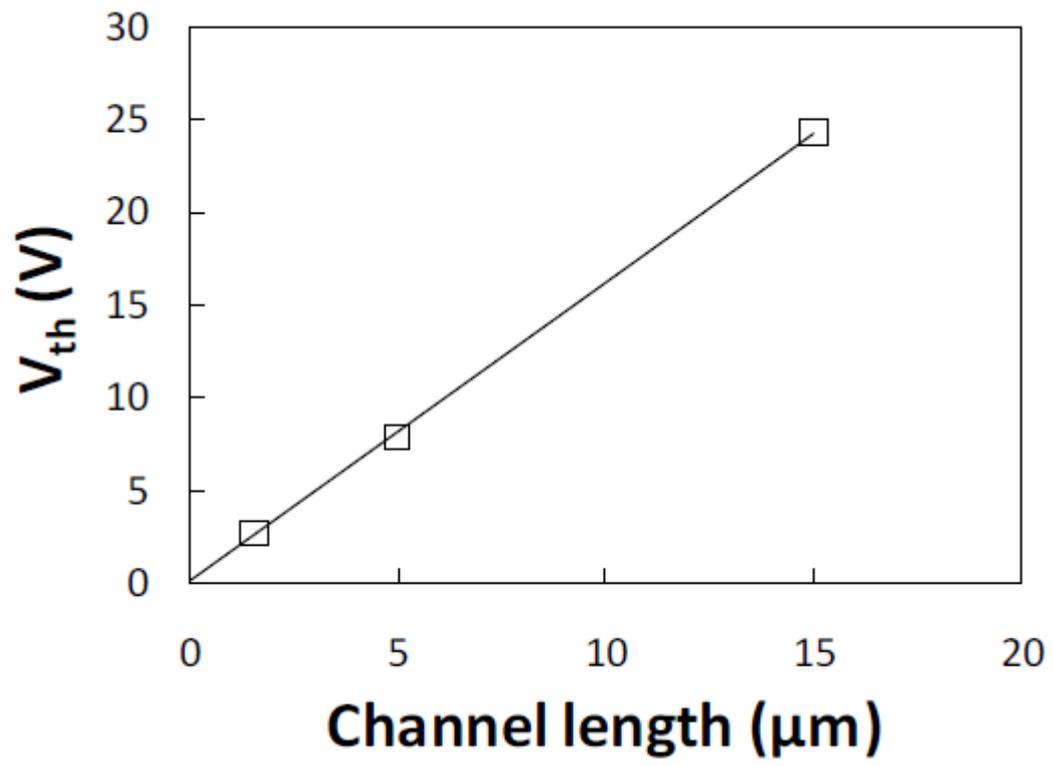

Fig. 1d Y. Sakai et al.

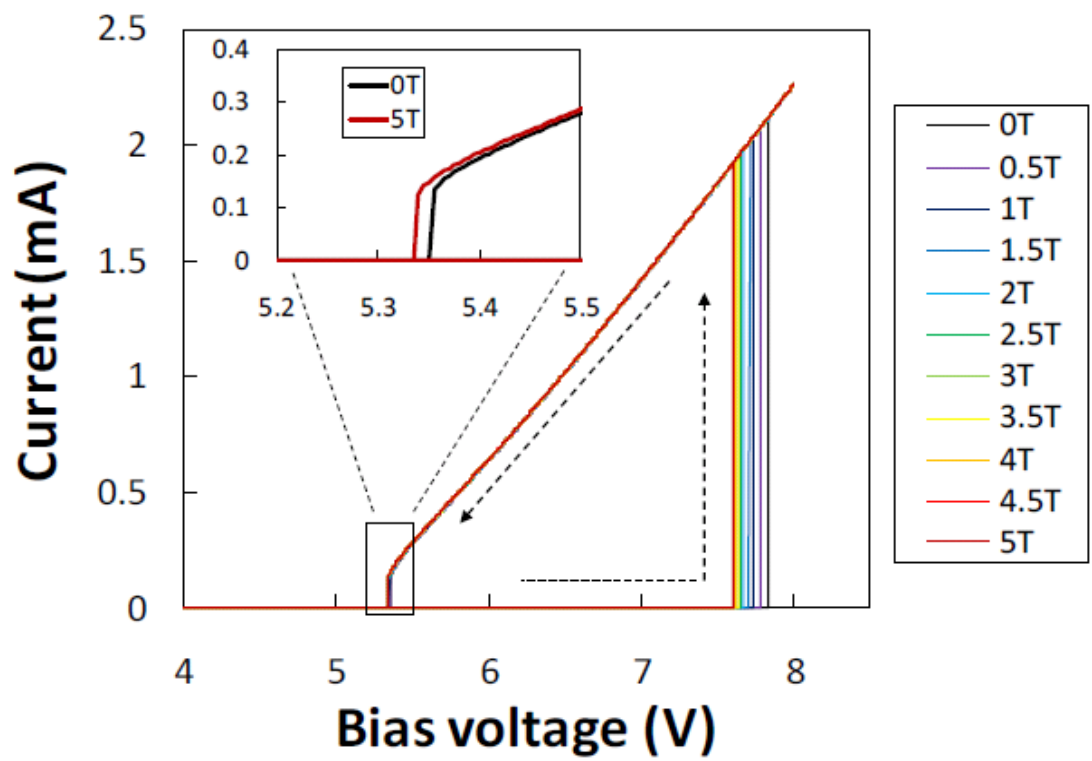

Fig. 2a Y. Sakai et al.

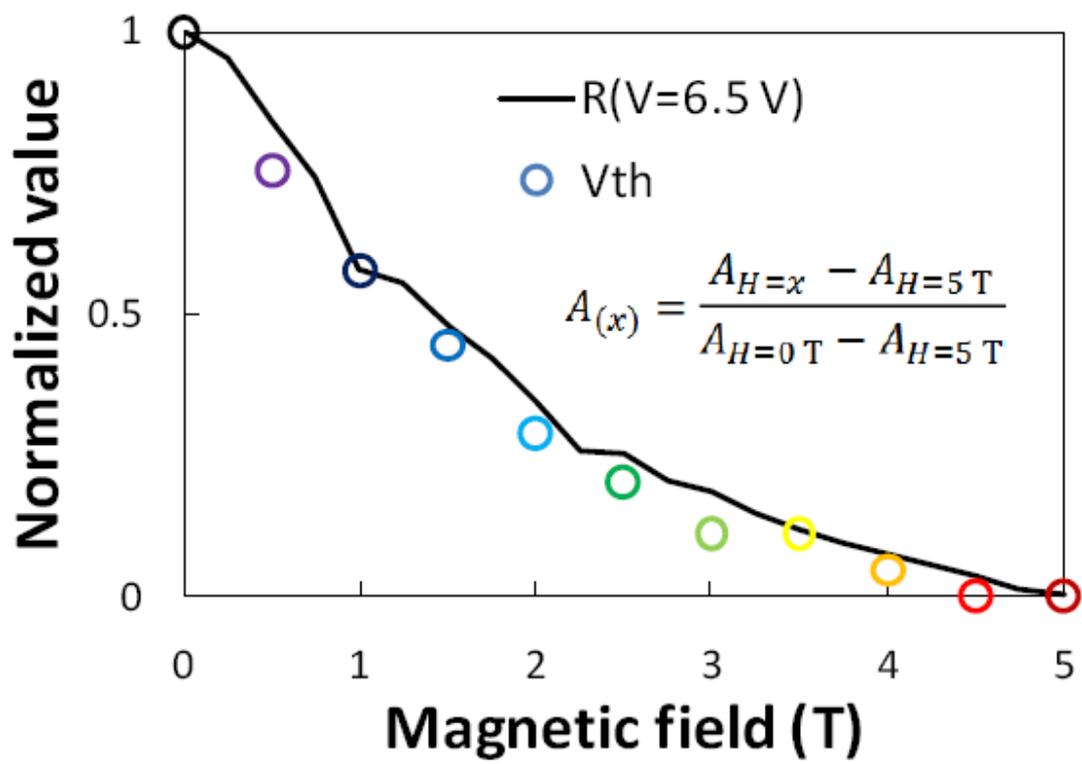

Fig. 2b Y. Sakai et al.

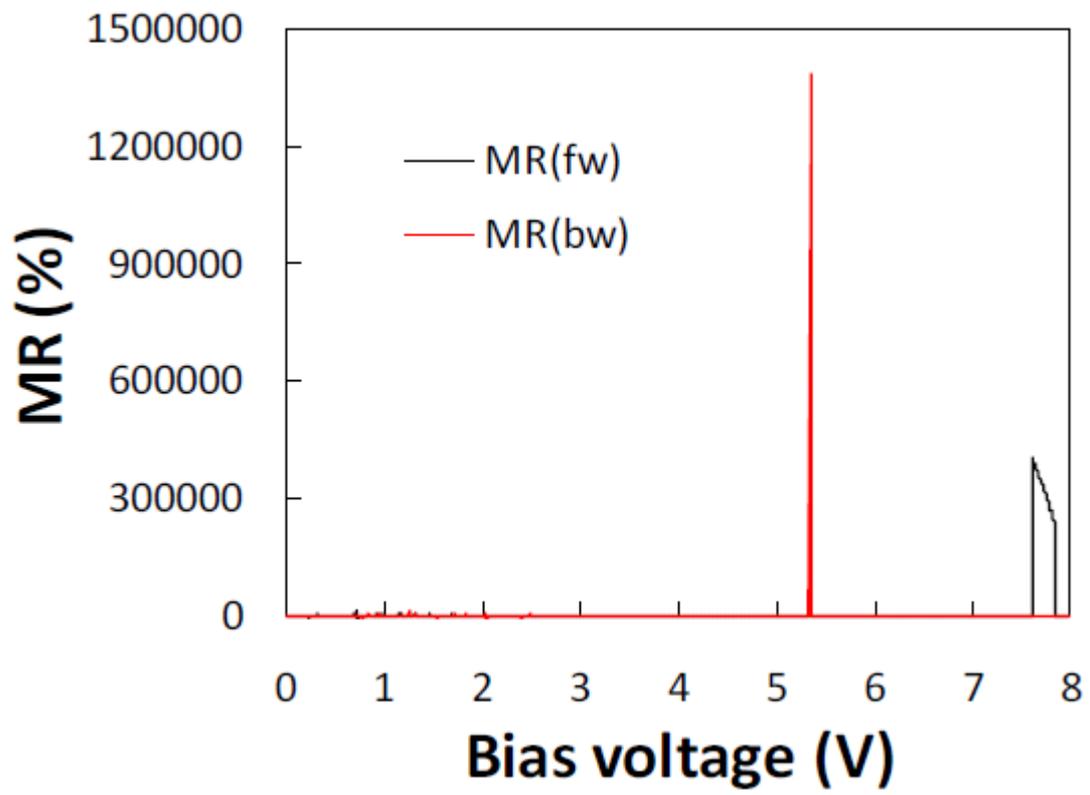

Fig. 2c Y. Sakai et al.

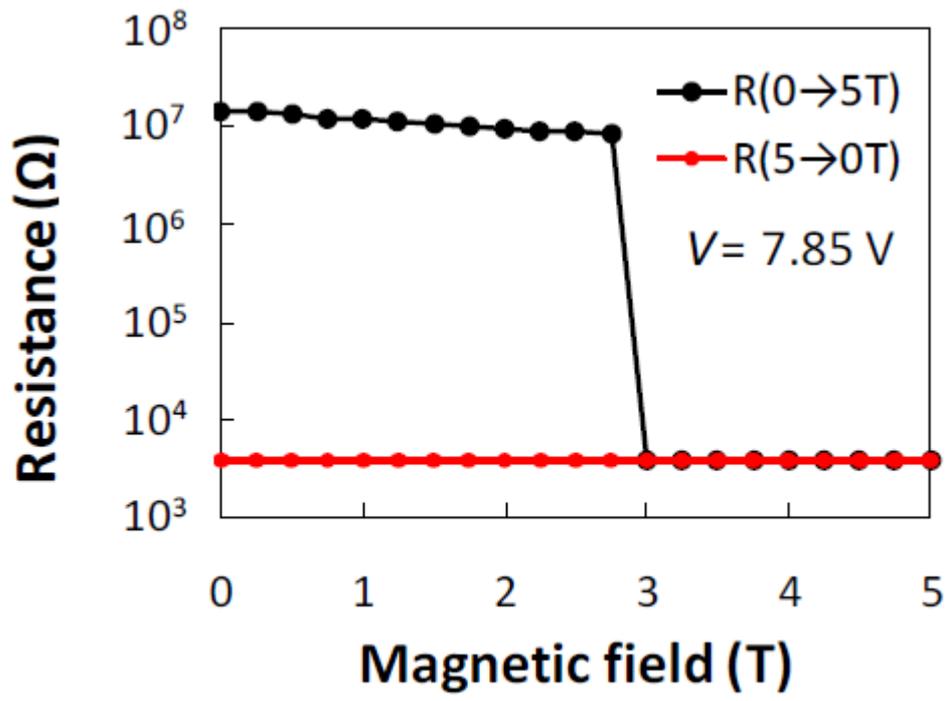

Fig. 3 Y. Sakai et al.

(a)

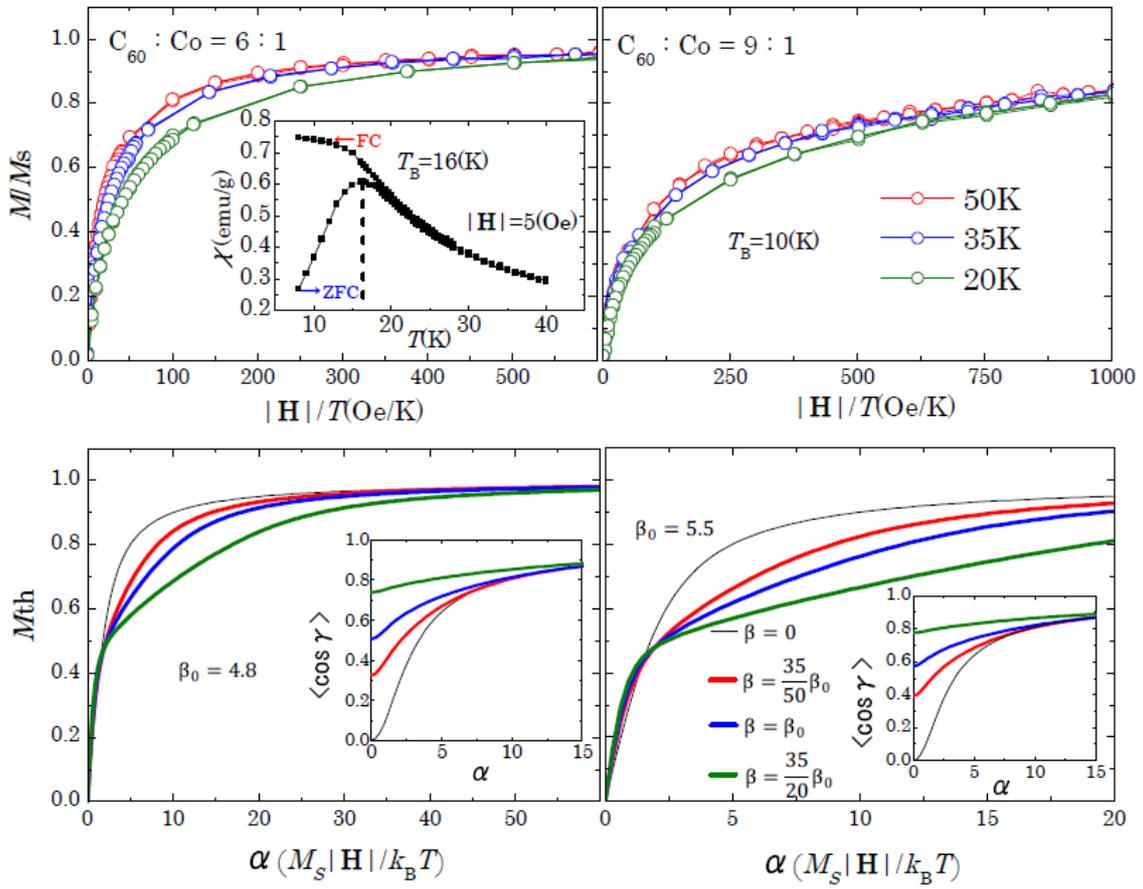

Fig. 4a Y. Sakai et al.

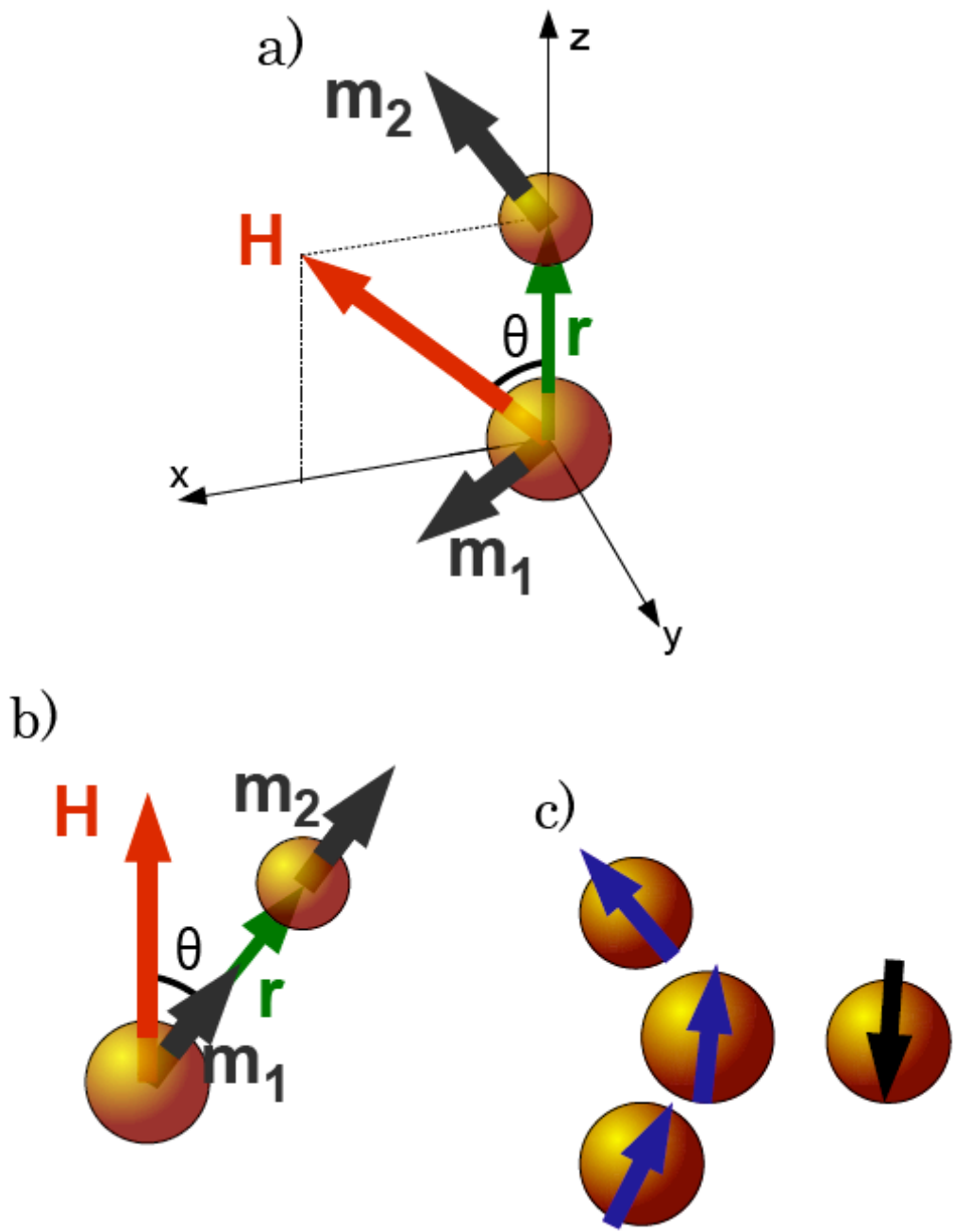

Fig. 5 Y. Sakai et al.

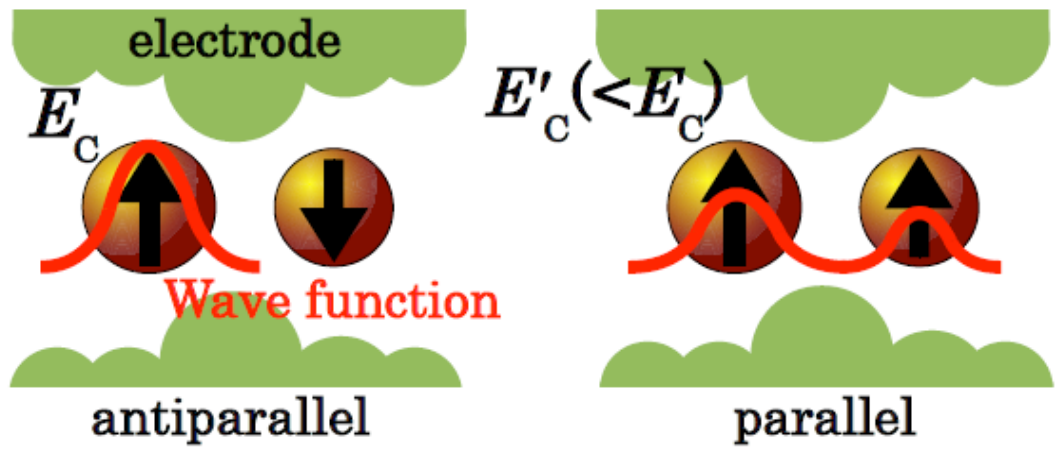

Fig. 6a Y. Sakai et al.

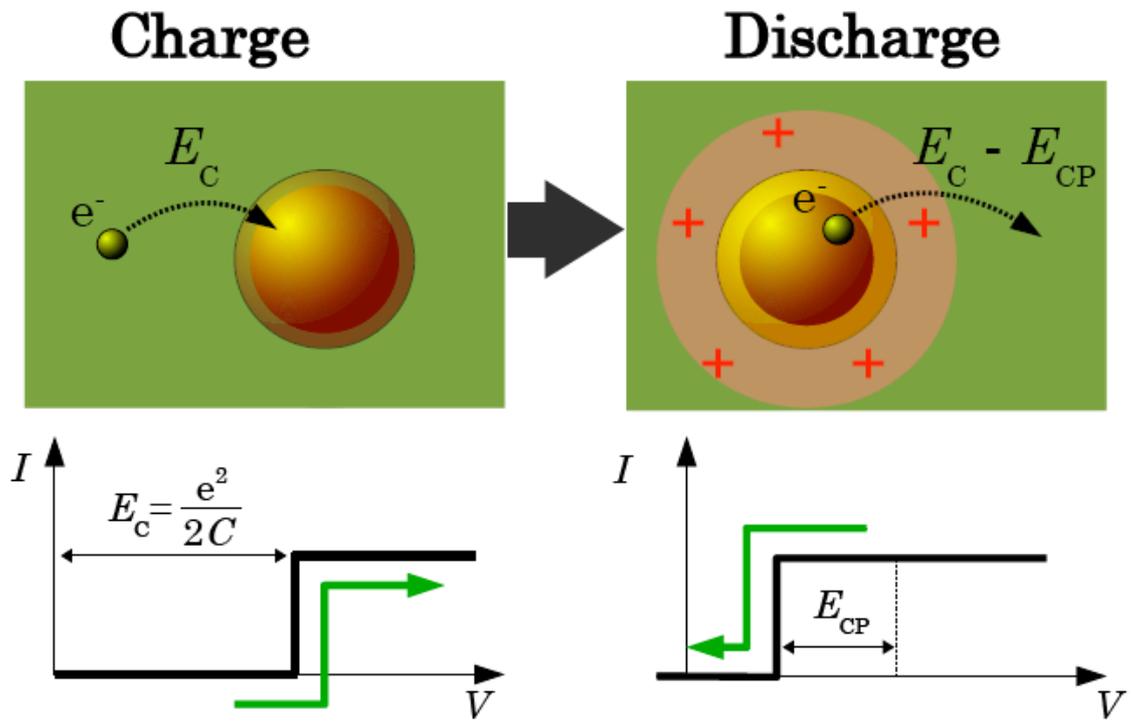

Fig. 6b Y. Sakai et al.

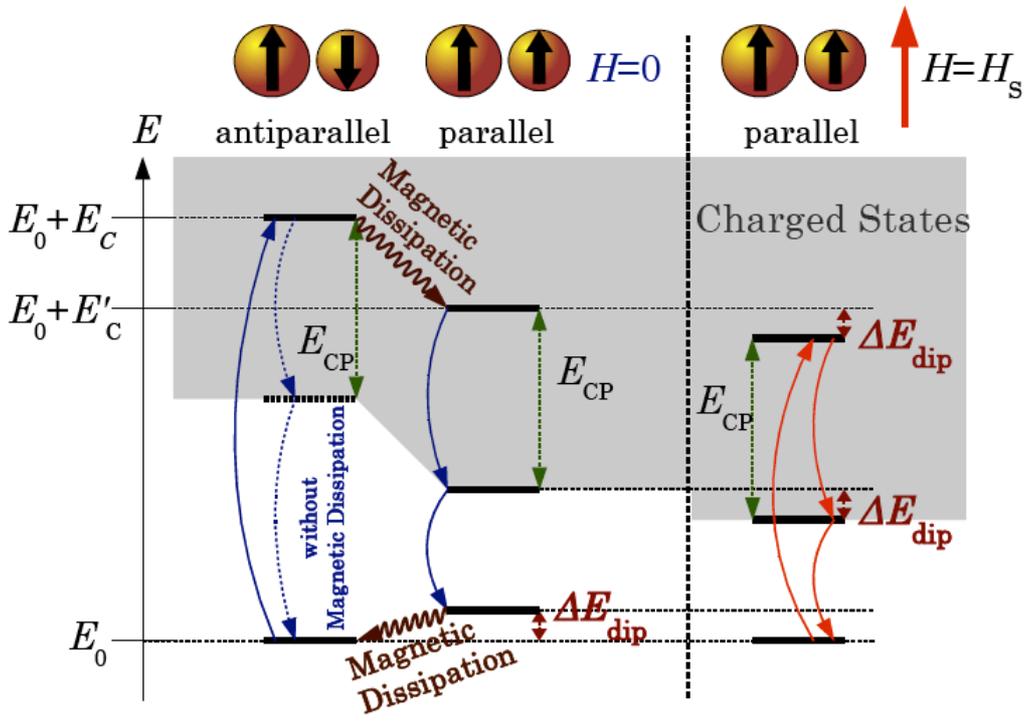

Fig. 6c Y. Sakai et al.